\documentclass[shortbibliography,twocolumn,prl,aps,superscriptaddress,showpacs,amsmath,amssymb,floatfix]{revtex4-2}
\usepackage{bm}
\usepackage{xcolor}
\usepackage{amsfonts}
\usepackage{amsmath}
\usepackage{amssymb}
\usepackage{graphicx}
\usepackage{tabularx}
\usepackage{multirow}
\usepackage{braket}
\usepackage{commath}%
\usepackage{comment}
\usepackage[colorlinks=true, letterpaper=true, pdfstartview=FitV,  linkcolor=blue, citecolor=blue, urlcolor=blue]{hyperref}



\begin{document}

\title{Crossed Electric Spin Generation in Multilayers}

\author{Xukun Feng}
\affiliation{State Key Laboratory of Surface Physics and Interdisciplinary Center for Theoretical Physics and Information Sciences, Fudan University, Shanghai 200433, China}

\author{Wei Du}
\affiliation{State Key Laboratory of Surface Physics and Interdisciplinary Center for Theoretical Physics and Information Sciences, Fudan University, Shanghai 200433, China}

\author{Jin Cao}
\affiliation{Research Laboratory for Quantum Materials, Department of Physics and Materials, The Hong Kong Polytechnic University, Hong Kong, China}

\author{Qian Niu}
\affiliation{CAS Key Laboratory of Strongly-Coupled Quantum Matter Physics, and Department of Physics,
University of Science and Technology of China, Hefei, Anhui 230026, China}

\author{Shengyuan A. Yang}
\email{shengyuan.yang@polyu.edu.hk}
\affiliation{Research Laboratory for Quantum Materials, Department of Physics and Materials, The Hong Kong Polytechnic University, Hong Kong, China}

\author{Cong Xiao}
\email{congxiao@fudan.edu.cn}
\affiliation{State Key Laboratory of Surface Physics and Interdisciplinary Center for Theoretical Physics and Information Sciences, Fudan University, Shanghai 200433, China}

\begin{abstract}
We propose a crossed electric spin generation effect in multilayer structures. This effect combines the advantages of in-plane and gate electric fields for manipulating spin, and unveils a nice quantum geometric structure in terms of the underlying hybrid Berry curvatures that are unique to quasi-2D systems and manifest inherent spin-layer-orbital coupling. Combining our theory with first-principles calculations, we predict sizable spin
generation in 2D ferromagnetic Fe$_3$GeTe$_2$ and altermagnetic V$_2$Te$_2$O. Our theory opens a new Berry-curvature route for all-electric spin manipulation in 2D materials.

\end{abstract}

\maketitle


Electric manipulation of spin has been a major task of spintronics, holding promise for next-generation information technology. In three-dimensional (3D) bulk materials, the spin polarization induced by applied electric field is strongly constrained by symmetry, e.g., it is forbidden in centrosymmetric crystals. Recently, the research has been increasingly focused on
quasi-2D multilayer systems, motivated by the flexibility in engineering their properties and by the need of device miniaturization \cite{Mak2017valley,Xu2017,Mak2019Probing}. Notably, such systems can be readily controlled by electric gating, a well-developed technique providing an in-situ, accurate, and efficient method for tuning electronic properties \cite{Xu2018gateBCD,Ma2019,Yu2020Valley,Gao2021Layer,Gao2023QM,Liu2024OME,chen2024crossed,li2024dynamical,Zhang2024gate-spin,Xu2025axion,Wang2025antiferromagnetic,Yu2025EHE,Ma2026bistable,Singh2026IPAHE}.
As for the phenomenon of electric generation of spin polarization \cite{Manchon2008,Garate2009,Franz2010,Kurebayashi2014,Freimuth2014,Zelezny2014,Jungwirth2016room,Manchon2019}, the gate electric field offers an additional handle to break (e.g., inversion) symmetry, alleviating the symmetry constraint.
Therefore, in a crossed field configuration, i.e., under both in-plane electric field and gate field,
spin polarization may be efficiently induced in a broad class of crystals. However, this promising possibility has not been investigated.

In this work, we propose such a crossed electric spin generation (CESG) effect, which is bilinear in the in-plane electric field $\bm E$ and the gate field $\mathcal E$ exerted on a multilayer.  According to the symmetry character under time-reversal ($\mathcal T$) operation,
one distinguishes a $\mathcal{T}$-odd CESG (which is restricted to magnetic systems) and a $\mathcal{T}$-even CESG (which may exist regardless of magnetism). We show that the effect greatly extends the symmetry allowed material classes of electric spin manipulation. For example, we find $\mathcal{T}$-odd CESG is universally present in all type-I magnetic point groups, and $\mathcal{T}$-even CESG has a component compatible with all magnetic point groups.

We formulate the microscopic theory for both $\mathcal{T}$-odd and $\mathcal{T}$-even CESG, and unveil their quantum geometric origin. While the $\mathcal{T}$-even CESG connects to a multilayer anomalous spin polarizability (ASP), the
$\mathcal{T}$-odd CESG originates from a set of \emph{hybrid} Berry curvatures. Previous studies on Berry-phase physics in condensed matters have mostly focused on momentum-space Berry curvature $\Omega_{\bm k\bm k}$. However, the notion of Berry curvature can be extended to any parameter space, and a hybrid Berry curvature  $\Omega_{\alpha\beta}$ in general
manifests a coupling between the degrees of freedom associated to parameters $\alpha$ and $\beta$.
We show that $\mathcal{T}$-odd CESG involves all three hybrid Berry curvatures $\Omega_{\bm h\bm k}$, $\Omega_{\mathcal E\bm k}$, and $\Omega_{\mathcal E\bm h}$ ($\bm h$ is a fictitious Zeeman field that is conjugate to spin)
in the extended $\bm k$-$\bm h$-$\mathcal E$ parameter space, which capture the interplay between orbital, spin, and layer degrees of freedom in a nice geometric description.

Combined with first-principles calculations, we apply our theory to two concrete 2D materials --- ferromagnetic monolayer Fe$_3$GeTe$_2$ and altermagnetic bilayer V$_2$Te$_2$O, both of which forbid linear electric spin generation but support a significant CESG. Notably, in altermagnets, CESG may lead to a staggered spin polarization, which is needed for rotating the N\'{e}el vector.
Our work discovers a new physical effect, unveils the importance of quantum geometry manifested in hybrid Berry curvatures inherent in layered materials, and opens a new route for all-electric spin manipulation in quasi-2D systems.

\emph{\textcolor{blue}{Formulation and hybrid Berry curvatures}.}
Consider a quasi-2D multilayer system. Its areal spin density can be expressed as
\begin{eqnarray}
\boldsymbol{S}=\int\left[  d\boldsymbol{k}\right]  f_{n\bm k}
\bm s_{n\bm k}, \label{spin density}
\end{eqnarray}
where $[d\bm k]$ is shorthand notation for $\sum_{n}d\bm k/(2\pi)^{2}$, $n$ is the band index,
$f$ is the distribution function, and $\bm s_{n\bm k}\equiv\langle W_{n\bm k}|\hat{\bm s}|W_{n\bm k}\rangle$ is the expectation value of spin operator $\hat{\bm s}$ in the electron wave packet
$|W_{n\bm k}\rangle$ (which is centered at the cell-periodic Bloch state $|u_{n\bm k}\rangle$).

In the crossed field configuration, the perturbations of in-plane field $\bm E$ and gate field $\mathcal E$ will alter the distribution function $f$ and the average spin $\bm s_{n\bm k}$ carried by the wave packet.
According to Ref. \cite{Dong2020}, $\bm s_{n\bm k}$ can be expressed in a compact form by introducing a fictitious Zeeman field $\bm h$ that couples to spin via a term $\bm h\cdot\hat{\bm s}$ in the Hamiltonian (set $e=\hbar=1$):
\begin{eqnarray}\label{ss}
  \bm s_{n\bm k}=\left(\partial_{\boldsymbol{h}}\tilde{\varepsilon
}+\tilde{\Omega}_{\boldsymbol{hk}}\cdot\boldsymbol{E}\right)\Big|_{\boldsymbol{h}=0},
\end{eqnarray}
where $\tilde\varepsilon$ is the wave packet energy, $\tilde{\Omega}_{\boldsymbol{hk}}$ is the hybrid Berry curvature in $k$-$h$ space \cite{Franz2010,Kurebayashi2014,Freimuth2014,Yuriy2017,Dong2020,xiao2022NLSOT,Xiao2023ASP,Tang2024SOT}, and we drop the state label $n\bm k$ on the right hand side. The tilde in $\tilde\varepsilon$ and $\tilde{\Omega}_{\boldsymbol{hk}}$ highlights that these quantities include gate-field corrections.
It is understood that the limit $\bm h=0$ should be taken at the end of calculation, and hereafter we shall omit the
$|_{\boldsymbol{h}=0}$ symbol.

The gate electric field $\mathcal{E}\hat{z}$ couples to the layer polarization operator $\hat{\varrho}\equiv d\mathcal J$ and adds a perturbation term
$\Delta H= \mathcal{E} \hat{\varrho}$
to the Hamiltonian. Here, $d$ is the lattice constant along $z$, and $\mathcal J$ is the layer-pseudospin matrix \cite{zheng2025layer,chen2024crossed,zheng2024interlayer}: For a system having $N$ layers, $\mathcal J$ is the $z$-component matrix of pseudospin $(N-1)/2$. For example, when $N=2$, $\mathcal J=\text{diag}\left( \frac{1}{2}, - \frac{1}{2} \right)$.

To linear order of gate field, the corrected band energy is given by $\tilde{\varepsilon}_{n}=\varepsilon
_{n}+\mathcal{E}\varrho^{n}$, where $\varepsilon$ is the unperturbed band energy and $\varrho^{n}$ is the intraband matrix element of layer polarization (for simple notations, we suppress $\bm k$ labels whenever there is no risk of confusion).
Meanwhile, the corrected $k$-$h$ hybrid Berry curvature reads
\begin{eqnarray}
\tilde{\Omega}_{h_ik_j}=\Omega_{h_{i}k_{j}}+ \left(\partial_{h_{i}}a_{j}-\partial_{k_{j}}\mathfrak{a}_{i}\right).
    \label{h-k}
\end{eqnarray}
Here, $\Omega_{h_{i}k_{j}}=\partial_{h_{i}}\mathcal{A}_{j}-\partial_{k_{j}}\mathfrak{A}_{i}$ is the one in the absence of gate field, with $\boldsymbol{\mathcal{A}}  =\langle u|i\partial_{\boldsymbol{k}}|u\rangle$
and $\bm{\mathfrak{A}}=\langle u|i\partial_{\bm h}|u\rangle$ being the
pertinent zero-field Berry connections. For a given band $n$, it is expressed as
\begin{equation}
\Omega_{h_{i}k_{j}}^{n}=\tilde{\Omega}^n_{h_ik_j}\big|_{\mathcal{E}=0}=-2{\operatorname{Im}}\sum_{\ell\neq
n}\frac{s_i^{n\ell} v_j^{\ell n}}{(\varepsilon
_{n}-\varepsilon_{\ell})^{2}},
\end{equation}
where $s^{n\ell}$ and $v^{\ell n}$ are the interband spin and velocity matrix elements.

The second term in Eq.~(\ref{h-k}) is the gate-field induced hybrid Berry curvature, where $\bm a$ ($\bm{\mathfrak{a}}$) is the induced $k$-space ($h$-space) Berry connection. Interestingly, their derivatives with respect to gate field, i.e.,  $\partial_{\mathcal{E}}\bm a$ and $\partial_{\mathcal{E}}\bm{\mathfrak{a}}$, correspond to another two hybrid Berry curvatures. To linear order in $\mathcal E$, we have
\begin{align}
\boldsymbol{a} & =\mathcal{E}{\Omega}_{\mathcal{E}\boldsymbol{k}}, \label{BCP-1}\\
\bm{\mathfrak{a}} & =\mathcal{E}{\Omega}_{\mathcal{E}\boldsymbol{h}}. \label{BCP-2}
\end{align}
Here, ${\Omega}_{\mathcal{E}\boldsymbol{k}}$ and ${\Omega}_{\mathcal{E}\boldsymbol{h}}$ are $\mathcal{E}$-$k$ and $\mathcal{E}$-$h$ hybrid Berry curvatures, respectively, which are defined similar to $\tilde{\Omega}_{hk}$.
For a specific band $n$, we have
\begin{align}
\Omega_{\mathcal{E}\boldsymbol{k}}^{n} & \equiv \tilde{\Omega}^n_{\mathcal{E}\boldsymbol{k}}\big|_{\mathcal{E}=0}=-2
{\operatorname{Im}}\sum_{\ell \neq n}\frac{\varrho^{n \ell}\boldsymbol{v}^{\ell n}}{(\varepsilon_{n}-\varepsilon_{\ell})^{2}}, \notag \\
\Omega_{\mathcal{E}\boldsymbol{h}}^{n} & \equiv \tilde{\Omega}^n_{\mathcal{E}\boldsymbol{h}}\big|_{\mathcal{E}=0}=-2
{\operatorname{Im}}\sum_{\ell\neq n}\frac{\varrho^{n\ell}\boldsymbol{s}^{\ell n}}{(\varepsilon_{n}-\varepsilon_{\ell})^{2}}.
\end{align}
Noticeably, $\Omega^n_{\mathcal{E}\boldsymbol{k}}$ and $\Omega^n_{\mathcal{E}\boldsymbol{h}}$ depend on $\varrho^{n\ell}$, the interband element of layer polarization operator, which vanishes if state $\ket{u_{n \bm k}}$ is completely polarized in one of the layers. Hence, these two Berry curvatures must require interlayer hybridization of electronic states.

Before proceeding, we elucidate the physical content of the involved hybrid Berry curvatures. Generally, $\Omega_{\alpha\beta}$ captures the response of $\bar\alpha$ to $\dot\beta$ (or equivalently, for $\bar\beta$ and $\dot\alpha$), where $\bar\alpha=\partial H/\partial \alpha$ is the physical quantity conjugate to $\alpha$, and $\dot\beta$ is the rate of change of $\beta$ parameter. For example, taking $\alpha$ as $\bm h$, then $\bar\alpha$ is the electron spin. Under in-plane driving field, $\dot{\bm k}=-\bm E$, so $\Omega_{\bm h\bm k}$ describes $\bm E$ field induced spin, which explains the second term in Eq.~(\ref{ss}). Extending this analysis to Berry curvatures in
the extended $\bm k$-$\bm h$-$\mathcal E$ parameter space, we can organize the equations of motion of an electron wave packet in a nice form:

\begin{align}
    \begin{bmatrix}
        \bm v \\ \bm s \\ \varrho
    \end{bmatrix}
    = \begin{bmatrix}
        \partial_{\bm k} \\ \partial_{\bm h} \\ \partial_{\mathcal{E}}
    \end{bmatrix} \varepsilon -
    \begin{bmatrix}
        \Omega_{\boldsymbol{kk}} & \Omega_{\boldsymbol{kh}} & \Omega_{\boldsymbol{k}\mathcal{E}} \\
         \Omega_{\boldsymbol{hk}} & \Omega_{\boldsymbol{hh}} & \Omega_{\boldsymbol{h}\mathcal{E}} \\
         \Omega_{\mathcal{E}\boldsymbol{k}} & \Omega_{\mathcal{E}\boldsymbol{h}}& \Omega_{\mathcal{E}\mathcal{E}}
    \end{bmatrix}
    \begin{bmatrix}
        \dot{\boldsymbol{k}} \\ \dot{\boldsymbol{h}} \\ \dot{\mathcal{E}}
    \end{bmatrix}.\label{99}
\end{align}
The second term contains an antisymmetric matrix (since $\Omega_{\alpha\beta}=-\Omega_{\beta\alpha}$) involving all six types of Berry curvatures in $\bm k$-$\bm h$-$\mathcal E$ space.
Rich quantum-geometric response phenomena can be read from this matrix. Besides $\Omega_{\boldsymbol{kk}}$ and $\Omega_{\boldsymbol{hh}}$ \cite{Xiao2010}, what is relatively familiar is $\Omega_{\boldsymbol{hk}}$, which underlies the in-plane $E$-field induced spin ($\Omega_{\boldsymbol{hk}}\dot{\boldsymbol{k}}$, $\dot{\boldsymbol{k}}=-\bm E$) \cite{Freimuth2014,Yuriy2017,Dong2020} and the adiabatic charge current driven by magnetic dynamics ($\Omega_{\boldsymbol{kh}}\dot{\boldsymbol{h}}$) \cite{Xiao2021SOT}. $\Omega_{\mathcal{E}\boldsymbol{k}}$ dictates in-plane $E$-field induced out-of-plane electric polarization ($\Omega_{\mathcal{E}\boldsymbol{k}}\dot{\boldsymbol{k}}$) \cite{zheng2024interlayer} as well as adiabatic current driven by ac gate field ($\Omega_{\boldsymbol{k}\mathcal{E}}\dot{\mathcal{E}}$). The latter phenomenon also implies an in-plane electric polarization induced by gate field ($\Omega_{\boldsymbol{k}\mathcal{E}}\mathcal{E}$), which is what we encounter in Eq.~(\ref{BCP-1}) (since $\bm a$ physically represents a positional shift of electron wave packet~\cite{Gao2014}). Therefore, $\Omega_{\mathcal{E} \boldsymbol{k}}$ represents a kind of crossed polarizability.
Finally, $\Omega_{\mathcal{E}\boldsymbol{h}}$ induces two novel dynamic magnetoelectric effects in multilayer systems: out-of-plane electric polarization induced by magnetic dynamics or time varying Zeeman field  ($\Omega_{\mathcal{E}\boldsymbol{h}}\dot{\boldsymbol{h}}$), and spin polarization driven by time varying gate field ($\Omega_{\boldsymbol{h}\mathcal{E}}\dot{\mathcal{E}}$). The latter response can also be formulated as $\bm s=-\tilde{\Omega}_{\bm h t}=\dot{\mathfrak{a}}$, with $\mathfrak{a}$ given in Eq.~(\ref{BCP-2}). As such, $\Omega_{\mathcal{E}\boldsymbol{h}}$ represents the dynamic spin magnetoeletric polarizability, or called dynamic ASP \cite{Feng2025DESG}, for multilayer systems.

\begin{table*}[t]
	\caption{\label{tablesymmetry}Symmetry constraints on the $\mathcal{T}$-odd CESG. The considered symmetry operations are compatible with a quasi-2D system in the $xy$ plane. $\checkmark$ ($\times$) denotes a symmetry-allowed (forbidden) component. The operations $\mathcal{T}$, $\mathcal{PT}$, $C_3^z\mathcal{T}$, $C_6^z\mathcal{T}$ and $S_6^z\mathcal{T}$ forbid all components and are not listed.}
	\begin{ruledtabular}
		\begin{tabular}{cccccccccccc}		
			Component & $\mathcal{P}$ & $C_{2}^{z}$ & $C_{3,4,6}^{z},\,S_{4,6}^{z}$ & $C_{2}^{x}$ & $\mathcal{M}_z$ & $\mathcal{M}_x$ & $C_{2}^{z}\mathcal{T}$ & $C_{4}^{z}\mathcal{T},\,S_{4}^{z}\mathcal{T}$ & $C_{2}^{x}\mathcal{T}$ & $\mathcal{M}_z\mathcal{T}$ & $\mathcal{M}_x\mathcal{T}$   \\
			\hline
            $\alpha_{xx}$ & $\checkmark$ & $\checkmark$ & $\checkmark$ & $\times$ & $\checkmark$ & $\times$ & $\times$ & $\checkmark$ & $\checkmark$ & $\times$ & $\checkmark$ \\
            $\alpha_{xy}$ & $\checkmark$ & $\checkmark$ & $\checkmark$ & $\checkmark$ & $\checkmark$ & $\checkmark$ & $\times$ & $\checkmark$ & $\times$ & $\times$ & $\times$ \\
            $\alpha_{yx}$ & $\checkmark$ & $\checkmark$ & $-\alpha_{xy}$ & $\checkmark$ & $\checkmark$ & $\checkmark$ & $\times$ & $\alpha_{xy}$ & $\times$ & $\times$ & $\times$ \\
            $\alpha_{yy}$ & $\checkmark$ & $\checkmark$ & $\alpha_{xx}$ & $\times$ & $\checkmark$ & $\times$ & $\times$ & $-\alpha_{xx}$ & $\checkmark$ & $\times$ & $\checkmark$ \\
            $\alpha_{zx}$ & $\checkmark$ & $\times$ & $\times$ & $\checkmark$ & $\times$ & $\checkmark$ & $\checkmark$ & $\times$ & $\times$ & $\checkmark$ & $\times$ \\
            $\alpha_{zy}$ & $\checkmark$ & $\times$ & $\times$ & $\times$ & $\times$ & $\times$ & $\checkmark$ & $\times$ & $\checkmark$ & $\checkmark$ & $\checkmark$ \\
		\end{tabular}
	\end{ruledtabular}
    \label{T-odd symmetry}
\end{table*}

\emph{\textcolor{blue}{$\mathcal T$-odd CESG}.}
Substituting the above hybrid Berry curvatures into Eq. (\ref{spin density}) and retaining to the $\mathcal T$-odd contributions $\propto E\mathcal{E}$, we obtain the $\mathcal T$-odd CESG:
\begin{equation}
\delta S^{\text{int}}_{i}=\alpha_{ij}^{\text{int}}E_{j}\mathcal{E},
\end{equation}
where the response coefficient is determined by the unperturbed (zero-field) band structure:
\begin{eqnarray}
    \alpha_{ij}^{\text{int}}=  \int\left[  d\boldsymbol{k}\right]
\left[\partial_{h_{i}}\Omega_{\mathcal{E}k_{j}}f_{0} + (
v_{j}\Omega_{\mathcal{E}h_{i}} +\varrho\Omega_{h_{i}k_{j}})f'_{0} \right],\label{key}
\end{eqnarray}
where $f_{0}(\varepsilon)$ is the Fermi-Dirac distribution function, $v$ and $\varrho$ denote the corresponding intraband matrix elements.

We have the following observations. First, one can easily check that all terms in (\ref{key}) are indeed $\mathcal T$-odd, so this response can exist only in magnetic multilayers. Second, $\alpha_{ij}^{\text{int}}$ represents an \emph{intrinsic} response, meaning that it is determined solely by the band structure, independent of scattering. Such intrinsic response coefficients are usually a focus of interest. Third, it is interesting that $\alpha_{ij}^{\text{int}}$ involves all three types of hybrid Berry curvatures of the $\bm k$-$\bm h$-$\mathcal E$ space, which manifest a coupling between orbital, spin, and layer degrees of freedom underlying this effect. This differs from many previously studied intrinsic responses, such as the anomalous Hall effect, where only momentum-space Berry curvature $\Omega_{\bm k\bm k}$ plays the major role.

In insulators, the Fermi surface term in Eq.~(\ref{key}) vanishes, and the response reduces to the \textit{h}-derivative of the crossed polarizability $\mathbf{\Omega}_{\mathcal{E}\boldsymbol{k}}$ integrated over all valence bands, i.e., $\alpha_{ij}^{\text{int}}=\partial_{h_{i}}\int \left[d\boldsymbol{k}\right]\Omega_{\mathcal{E}k_{j}}$. This result has a simple thermodynamic interpretation: the polarization energy due to $\bm E$ and $\mathcal E$ fields is given by $\int \left[d\boldsymbol{k}\right]\Omega_{\mathcal{E}\bm k}\cdot \bm E \mathcal{E}$, so its $h$-derivative yields the induced spin polarization. We also comment that factors like
$\partial_{h_{i}}\mathbf{\Omega}_{\mathcal{E}\boldsymbol{k}}$ and other Berry curvatures in (\ref{key}) can be expressed in a form involving only unperturbed band matrix elements~\cite{supp}, which are amenable to first-principles calculations.

\emph{\textcolor{blue}{$\mathcal T$-even CESG}.}
By collecting $\mathcal T$-even contributions $\propto E\mathcal{E}$ in Eq.~(\ref{spin density}), one obtains the $\mathcal T$-even CESG. One finds this effect must involve off-equilibrium electrons, hence it is of extrinsic character.
Again, defining the response coefficient via $\delta S_{i}^{\text{eve}}=\alpha_{ij}^{\text{eve}}E_{j}\mathcal{E}$, we find
\begin{eqnarray}
\alpha_{ij}^{\text{eve}}=\tau\int\left[  d\boldsymbol{k}\right] \left( v_{j}\mathcal{Y}_i +  \varrho\partial_{k_{j}}s_{i} \right) f_{0}^{\prime}, \label{even}
\end{eqnarray}
where $\tau$ is the relaxation time, and
\begin{equation}
\mathcal{Y}_i^n=\partial_{\mathcal{E}}\tilde{s}_i^n\big|_{\mathcal{E}=0}
=-2{\operatorname{Re}}\sum_{\ell\neq n}\frac{\varrho^{n\ell}s_{i}^{\ell n}%
}{\varepsilon_{n}-\varepsilon_{\ell}}
\label{layer ASP}
\end{equation}
is the ASP for multilayer systems, which measures the anomalous spin polarization induced by gate field. Compared to the usual ASP defined in 3D systems, which is essentially the hybrid Berry
curvature $\Omega_{hk}$~\cite{Xiao2023ASP}, $\mathcal{\bm Y}$ is not of the form of a Berry curvature, and it requires
interlayer hybridization (as embodied in $\varrho^{n\ell}$ factor).
The first and second terms of Eq.~(\ref{even}) are favored by layer-hybridized and layer-polarized Bloch states on the Fermi surface, respectively.
In addition, one can see this effect is a Fermi surface response, so it is suppressed in insulators.

\emph{\textcolor{blue}{Symmetry analysis.}} Under a magnetic point group operation,
the intrinsic CESG $\alpha_{ij}^{\mathrm{int}}$ transforms as the component $\alpha_{ijz}^{\mathrm{int}}$ of a rank-3 $\mathcal{T}$-odd pseudotensor:
\begin{equation}
    \alpha_{i'j'}^{\mathrm{int}}=\eta_T \det(O)O_{i'i}O_{j'j}O_{zz}\alpha_{ij}^{\mathrm{int}},
\end{equation}
where $O$ denotes the spatial part of the operation, $\eta_T=+1$ for an non-primed (i.e., not containing $\mathcal T$) operation and $\eta_T=-1$ for a primed operation.
For a quasi-2D system in the $xy$ plane, $O$ must fulfills $O\hat{ z}=\pm\hat{ z}$.
The constraints imposed by symmetries are summarized in Table~\ref{T-odd symmetry}.

Notably, inversion symmetry $\mathcal{P}$ is compatible with CESG.
In contrast, $\mathcal{T}$ and $\mathcal{PT}$ forbid $\alpha_{ij}^{\mathrm{int}}$ because of its $\mathcal{T}$-odd character.
Importantly, for the in-plane spin response, i.e., for $i,j \in \{x,y\}$, we find the antisymmetric part of $\alpha_{ij}^{\mathrm{int}}$ transforms as a $\mathcal{T}$-odd \emph{scalar}, indicating that it is universally allowed in all type-I magnetic point groups.
On the other hand, the $(i,j)$ symmetric components of $\alpha_{ij}^{\mathrm{int}}$ are permitted by horizontal mirror ${M}_z$ and twofold rotation ${C}_2^z$, but are forbidden by ${M}_z \mathcal{T}$ and ${C}_{2}^z \mathcal{T}$.

As for $\mathcal{T}$-even CESG, $\alpha_{ij}^{\mathrm{eve}}$ transforms as the component $\alpha_{ijz}^{\mathrm{eve}}$ of a rank-3 $\mathcal{T}$-even pseudotensor:
\begin{equation}
\alpha_{i'j'}^{\mathrm{eve}}=\det(O)O_{i'i}O_{j'j}O_{zz}\alpha_{ij}^{\mathrm{eve}}.
\end{equation}
Its symmetry constraints correspond to the columns of non-primed operations in Table~\ref{T-odd symmetry}. For the in-plane spin response, the antisymmetric part of $\alpha_{ij}^{\mathrm{eve}}$ transforms as a $\mathcal{T}$-even \emph{scalar}, and hence has no limitations. Particularly, we find that in nonmagnetic dihedral point groups $D_n$ and achiral polar groups $C_{nv}$ ($n=3,4,6$), only the $(x,y)$-antisymmetric component of $\alpha_{ij}^{\mathrm{eve}}$ is allowed.

\begin{figure}[pt]
\includegraphics[width=8.5cm]{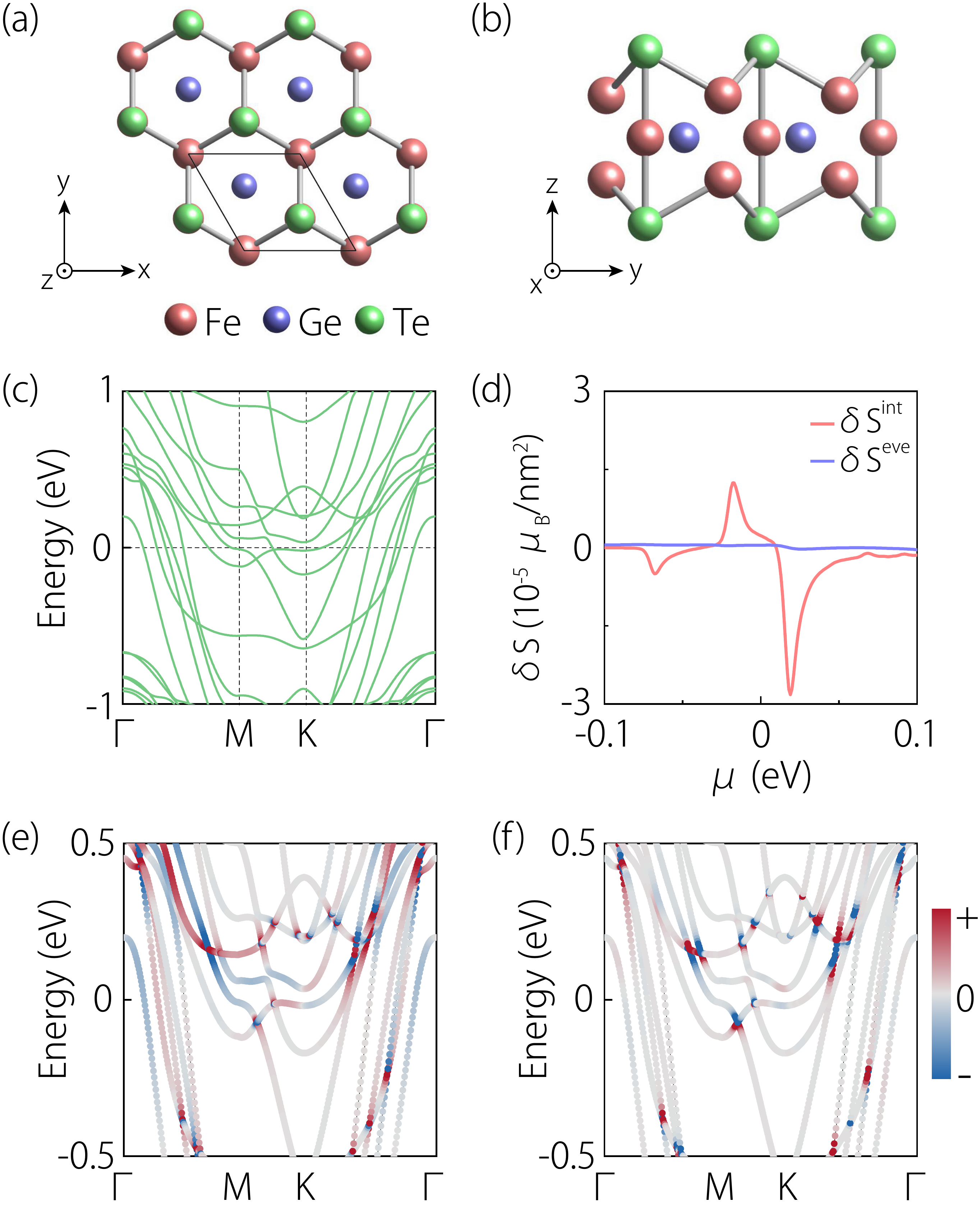}
\caption{(a) Top and (b) Side view of monolayer Fe$_3$GeTe$_2$. (c) Calculated band structure (SOC included). (d) Calculated CESG $\delta S^{\text{int}}$ and $\delta S^{\text{eve}}$ versus chemical potential, with $E=1$ kV/cm and $\mathcal{E}=0.01$ V/nm. (e-f) Distribution of hybrid Berry curvatures (e) $\Omega_{\mathcal{E}h_x}$ and (f) $\Omega_{\mathcal{E}k_x}$ in the band structure. In the calculation, the temperature is set to 50 K.}
\label{figFGT}
\end{figure}

\emph{\textcolor{blue}{Application to monolayer \text{Fe$_3$GeTe$_2$}.}}
We first apply our theory to monolayer Fe$_3$GeTe$_2$, an experimentally realized 2D ferromagnet with a Curie temperature $\sim 75$ K~\cite{fei2018two,Deng2018Gate}.
Monolayer Fe$_3$GeTe$_2$ has a hexagonal lattice with point group $D_{3h}$ [Figs.~\ref{figFGT}(a) and \ref{figFGT}(b)].
Its out-of-plane magnetic easy axis leads to the magnetic point group $\bar{6}m'2'$~\cite{fei2018two,shen2021magnetic}, generated by $\mathcal{P}$, $C_3^z$, and $C_2^x\mathcal{T}$.
Importantly, these symmetries forbid all components of the linear spin response, but allow CESG.
According to Table~\ref{T-odd symmetry}, the nonzero components of $\alpha^{\mathrm{int}}$ are $\alpha^{\mathrm{int}}_{xx}=\alpha^{\mathrm{int}}_{yy}$, thus the $\mathcal{T}$-odd crossed response is forced to be aligned with the in-plane $E$ field. In comparison, the allowed components of $\alpha^{\mathrm{eve}}$ are $\alpha^{\mathrm{eve}}_{xy}=-\alpha^{\mathrm{eve}}_{yx}$ and so the $\mathcal{T}$-even crossed response must be transverse to both driving fields. Thus, we have
\begin{align}
    \delta \bm S^{\mathrm{int}} =\alpha^{\mathrm{int}}_{xx} \mathcal{E}\bm E, \quad \delta \bm S^{\mathrm{eve}} =\alpha^{\mathrm{eve}}_{yx} \mathcal{E}\hat{z}\times \bm E.
\end{align}
It is nice that both responses have their magnitudes independent of the in-plane direction of $\bm E$, and they are mutually orthogonal.

The calculated band structure of Fe$_3$GeTe$_2$ (see \cite{supp} for calculation methods) are displayed in Fig.~\ref{figFGT}(c), which exhibits several band near-degeneracies around Fermi level. Pronounced Berry curvatures are expected to appear across these regions. Figures~\ref{figFGT}(e-f) show the $k$-space distributions of the hybrid Berry curvatures $\Omega_{\mathcal{E}h_x}$ and $\Omega_{\mathcal{E}k_x}$. Their hot spots can indeed be traced to the small-gap regions.
We then evaluate $\alpha_{xx}^{\mathrm{int}}$ and estimate the induced $\delta S^{\mathrm{int}}$ for $E=1$ kV/cm and $\mathcal{E}=0.01$ V/nm (the field values are within experimentally accessible regimes \cite{Geller2009,Jungwirth2018,Xu2018gateBCD,Song2019,Gao2021Layer}).
As shown in Fig.~\ref{figFGT}(d), at the intrinsic Fermi level, $\delta S^{\mathrm{int}}$ reaches $2.3\times10^{-6}~\mu_B/\mathrm{nm}^2$ and is further enhanced to $\sim 3\times10^{-5}~\mu_B/\mathrm{nm}^2$ upon slight electron doping to $\mu=20$ meV. This is a significant value, when compared to previously measured linear spin generation in noncentrosymmetric magnets (where a spin polarization on the order of $\sim 10^{-8}~\mu_B/\mathrm{nm}^2$ can be detected~\cite{Kurebayashi2014,Manchon2019}).
We also evaluate the $\mathcal{T}$-even response $\alpha_{yx}^{\mathrm{eve}}$ (details in Ref. \cite{supp}).
Taking a typical $\tau\sim 50$ fs for 2D materials \cite{Ma2019} gives $\delta S^{\mathrm{eve}}\sim 4.7\times10^{-7}~\mu_B/\mathrm{nm}^2$ at the intrinsic Fermi level, which is two orders smaller than the intrinsic contribution.

\emph{\textcolor{blue}{Application to bilayer \text{V$_2$Te$_2$O}}.}
Another particularly appealing platform for studying our proposed effect is provided by 2D altermagnets. Such systems often preserve inversion symmetry $\mathcal{P}$, so the linear spin response is forbidden, but CESG could be supported. Moreover, CESG may generate \emph{staggered} spin polarization on the two magnetic sublattices, leading to a N\'eel spin-orbit torque, which can help to switch the N\'{e}el order~\cite{Zelezny2014}.

\begin{figure}[pt]
\includegraphics[width=8.5cm]{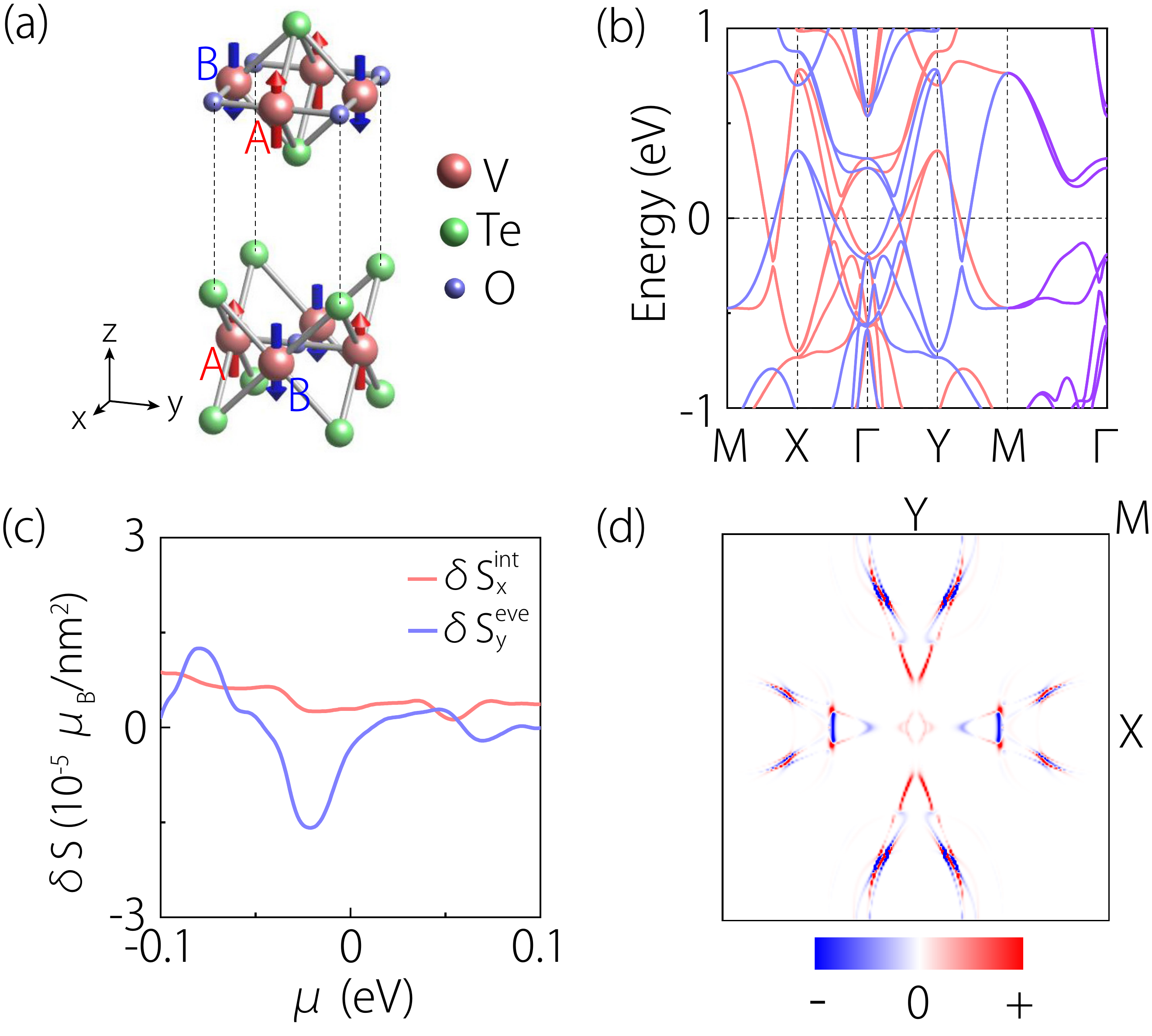}
\caption{(a) Crystal structure and magnetic ground state of bilayer V$_2$Te$_2$O. The sublattices A and B are marked. (b) Calculated band structure of bilayer V$_2$Te$_2$O without SOC. (c) Calculated staggered spin generation $\delta S_x^\text{int}$ and $\delta S_x^\text{eve}$ versus chemical potential, with $E=10^5$ V/m and $\mathcal{E}=0.01$ V/nm. The temperature is set to 50 K. (d) $k$-space distribution of the integrand in Eq.~(\ref{even}) for sublattice A.}
\label{figVTO}
\end{figure}

We demonstrate this mechanism in bilayer altermagnet V$_2$Te$_2$O~\cite{cui2023giant,liu2025anomalous}.
This material has a tetragonal lattice with crystallographic point group $D_{4h}$ [Fig.~\ref{figVTO}(a)].
Using first-principles calculations (details in \cite{supp}), its ground state magnetic configuration
can be identified, as shown in Fig.~\ref{figVTO}(a).
In the absence of spin-orbit coupling, its spin group is ${}^24/^1m^2m^1m$, exhibiting $d$-wave altermagnetism, with the characteristic $k$-dependent spin splitting evident in the calculated band structure [Fig.~\ref{figVTO}(b)].
Including spin-orbit coupling, the magnetic point group is $4'/mm'm$, generated by $\mathcal{P}$, $C_2^{x}\mathcal{T}$, $C_2^z$, and $C_4^z\mathcal{T}$.
We label the two magnetic sublattices as A and B (see Fig.~\ref{figVTO}(a)).
The magnetic site group of a single sublattice is $m'm'm$, which is generated by $C_2^x\mathcal{T}$, $C_2^z$ and $\mathcal{P}$. As mentioned, the presence of $\mathcal P$ forbids linear spin responses. As for CESG,
we find the allowed $\mathcal{T}$-odd responses on sublattice A are $\alpha_{xx}^\text{int,A}$ and $\alpha_{yy}^\text{int,A}$. Meanwhile, the $C_4^z\mathcal{T}$ symmetry which connects two sublattices imposes the constraint $\alpha_{xx}^{\text{int,A}}=-\alpha_{yy}^\text{int,B}$. Thus, a staggered spin generation is allowed. Meanwhile, the allowed components of $\mathcal{T}$-even CESG are $\alpha_{xy}^\text{eve,A}$ and $\alpha_{yx}^\text{eve,A}$, and it may also generate a staggered spin polarization.

We calculate the staggered spin generation for $\bm E$ applied along the $x$ direction. Taking $E=10^5$ V/m and $\mathcal{E}=0.01$ V/nm, the resulting staggered spin polarization versus chemical potential is plotted in Fig.~\ref{figVTO}(c). One finds that $\delta S^\text{int}_x\sim 2.9\times 10^{-6}$~$\mu_B/$nm$^2$ at the intrinsic Fermi level, and increases to $6.3\times 10^{-6}$~$\mu_B/$nm$^2$ upon hole doping to $\mu=-50$ meV. In comparison, the $\mathcal{T}$-even response gives $\delta S_y^\text{eve}\sim -3.3\times 10^{-6}$~$\mu_B/$nm$^2$ at intrinsic Fermi level and is further enhanced to $-1.6\times 10^{-5}$~$\mu_B/$nm$^2$ at $\mu=-20$ meV. These values are significant and can be readily detectable in experiment. In Fig.~\ref{figVTO}(d), we plot the $k$-space distribution of the integrand in the $\mathcal T$-even response
(\ref{even}), which is also enhanced around small-gap regions.

\emph{\textcolor{blue}{Discussion}.}
We have proposed a new physical effect of CESG. It can help to achieve electric manipulation of spin in a broad class of multilayer materials where the linear spin generation is symmetry forbidden.
We unveil its quantum geometric origin in a set of hybrid Berry curvatures characterizing the inherent orbital-spin-layer coupling of multilayer systems. The quasi-2D geometry and the gate control also make the effect highly relevant to device applications. Our predicted sizable spin polarization in concrete material examples can be readily detected by magneto-optical or anisotropic magnetoresistance experiments \cite{Manchon2019}, and may drive magnetization dynamics.
This opens new possibilities for spintronics.

We demonstrated CESG in two magnetic 2D materials. It should be noted that CESG, more specifically the $\mathcal T$-even CESG, also occurs in nonmagnetic materials. The resulting spin polarization may be used to drive an adjacent magnetic layer. And the effect offers a new mechanism for realizing a reconfigurable spin source.

The developed theory also resolves the crossed electric response of general local observables $\hat{\theta}$ in layered systems, as long as the $h$ field is replaced by the field conjugate to $\hat{\theta}$, which amounts to replacing $\hat{s}$ by $\hat{\theta}$ in the expressions of all the hybrid Berry curvatures and their derivatives.
Moreover, our remarks regarding hybrid Berry curvature and the structure exhibited in Eq.~(\ref{99}) is quite general. For a set of $N$ observables, in their equations of motion for a wave packet, there will be an $N\times N$ Berry curvature matrix. The off-diagonal entries of this matrix, i.e., the hybrid Berry curvatures, capture the cross coupling between different degrees of freedom. They will play an essential role towards quantum-geometry enabled multi-field control of condensed matters with entangled multiple degrees of freedom.

\bibliography{ref}

\end{document}